# Observation of Yamaji magic angles in bismuth surfaces.

Tito E. Huber[1*], Scott Johnson[1], Leonid Konopko[2,3] and Albina Nikolaeva[2,3]

[1] Howard University, Washington, DC 20059, USA

[2] Academy of Sciences, Chisinau, Moldova, MD-2028

[3] International Laboratory of High Magnetic Fields and Low Temperatures, 53-421 Wroclaw, Poland

Correspondence and requests for materials should be addressed to T.H. (email: thuber@howard.edu)

ABSTRACT

Bismuth consist of bismuth bilayers that are two-dimensional topological insulators and correspondingly, the surface is an array of edge states. Moreover, topological models, including second order topologic order, predict an interlayer electrical coupling mediated by hinge states. Here we report that angle dependent magnetoresistance measurements of small diameter single-crystal bismuth nanowires exhibit the sequence of magnetoresistance (MR) peaks at Yamaji magic angles and a peak for B//bilayer, indicating coherent transport between layers, and showing that the Fermi surface of surface electrons is a warped cylinder. The MR peaks are associated with magnetic field induced flat bands that are reminiscent of the well-known flat bands in bilayer graphene. Coherent transport across layers is interpreted in term of transport by topological hinge states.

Bi has been central in the research of topological insulators [1-8] because of its large spin-orbit coupling. Bi was believed to be a trivial topological insulator but later analysis has revealed it to be a topological semimetal[9-12]. The surface band is spin split at the M-point touching both conduction and valence band. The Bi crystal is layered and consists of a stack of bilayers, where the bilayers are two-dimensional TIs [8] and have one pair of protected helical modes on each edge. The individual Bi bilayers have not been isolated by exfoliation, but the bilayer that terminates the surface of bulk bismuth is sufficiently isolated that its study has provided a wealth of information. In this experiment, one-dimensional modes localized along edges, which were interpreted in terms of helical mode, as well as the two-dimensional (2D) surface states of Bi bilayers were identified [13,14]. Moreover, the crystal faceting of bismuth also may play an important role in bismuth. Recently, it has been proposed that bismuth is the first example of a conductor that exhibits a high order topology effect where its hinges, that contribute to the interlayer coupling, host topologically protected modes [15,16]. This is because different facets of a bismuth crystal have gaps of opposite sign and when such facets meet at hinges, the gap must pass through zero, producing a gapless hinge state. Electronic transport via the topological states in layers and facets, can explain the observations of extraordinary electronic mobility in bismuth thin films, nanowires, and ribbons [17-24]. However, there are few experimental probes that access the physics of layers and interlayer coupling. If the layered structure of the surface of bismuth was studied experimentally then one can examine the surface states, determine whether the layers possess topological modes and characterize the interlayer coupling. This can lead to future explorations of the physics of topological solids as layered materials as well as spintronic applications.

One method that has been successful in examining layered structures is the angular dependence of the magnetoresistance (AMR). When the magnetic field is rotated the resistance shows oscillations that are periodic in $\tan a$, $a$ being the angle between the magnetic field and the direction normal to the two-dimensional plane. These oscillations are different from the SdH oscillations in that they occur as a function of angle, but not a function of field intensity. This phenomenon was first recognized by Yamaji [25] in layered quasi-two dimensional organic conductors and it has since been observed in many layered materials [26], i.e. in non-magnetic layered oxides like $PdCoO_2$ [27]. Yamaji magic angles can appear when there is incoherent tunneling between layers [26]. The semiclassical theoretical theory by Yagi et al. shows that for coherent transport between layers, the oscillations arise at the Yamaji angles because the group velocity average (drift velocity), vanishes and the conductivity is zero [28]. Yamaji oscillations together with coherent interlayer transport indicate that the layered material Fermi surface is a warped and corrugated cylinder, an open Fermi surface (FS) [29-30]. Here we report the observation of Yamaji magic angles in the AMR of small diameter nanowires whose conduction is dominated by surface states and interlayer transport is coherent. The identification of the type of Fermi surface is key to our understanding of the metallic state and also the magnetoresistance, as open orbits in open Fermi surfaces are associated with non-saturating magnetoresistance [31]. The MR peaks are associated with magnetic field induced flat bands that are reminiscent of the well known flat bands in bilayer graphene and are known to promote correlated electron behavior[32,33].

**Results**

The diameters of the samples of single-crystal Bi nanowires used in our experiment ranged between 45 nm and 55 nm. The surface-to-volume ratio of our nanowires is high, and, therefore, surface effects are strongly expressed. Conductance studies of small-diameter Bi nanowires [10,11] and nanoribbons [11] clearly reveal surface conduction. Also, quantum confinement reduces the bulk carrier density thereby further increasing the contribution of the surface bands to nanowire electronic transport. Magnetoresistance and thermopower results for our Bi nanowires were presented previously [7]. We reported the observation of quantized conductance in the surface bands. With increasing magnetic fields oriented along the wire axis, the wires exhibit a stepwise increase in conductance and oscillatory thermopower, owing to an increased number of high-mobility spiral or helical surface modes. These modes appear with an $h/e$ and $h/2e$ periodicity, evidence of the 2D character of the surface. Our mobility estimates are comparable, within order of magnitude, to the mobility values reported for suspended graphene. The present work is based on a study of the transverse magnetoresistance.

We characterized the temperature dependence of the resistance and the dependence of the magnetoresistance with the orientation of the nanowires with the magnetic field. We observed a thermally activated conductivity that is typical of semiconductors at $T > 100$ K. At low temperatures, $T < 10$ K, the conductivity becomes saturated, indicating electronic transport by surface carriers.

We characterized the surface via a study of the Shubnikov-de Haas (SdH) oscillations (periodic in $1/B$) of the transverse magnetoresistance (TMR), where the field is perpendicular to the axis of the wire. We assigned the SdH oscillations to 2D Landau states of the surface carriers. Analysis of the temperature and magnetic field dependence of the SdH oscillations, produced $m_\Sigma = 0.25 \pm 0.03$ in units of the electron mass $m_0$, in good agreement with ARPES measurements

[34] The charge density per unit area $\Sigma$ was estimated from the SdH period ($P = 0.060\ T^{-1}$) using $\Sigma = f/(P\Phi_0)$, where $f$ is the 2D Landau level degeneracy, which is two on account of the two-fold spin degeneracy. We found $\Sigma = 8.06 \times 10^{11}/cm^2$, which was an order of magnitude smaller that the ARPES measurement of $8 \times 10^{12}/cm^2$ for Bi crystals. The 2D Fermi energy $E_F = \pi\hbar^2\Sigma/m_\Sigma$ and $k_F$ were found to be 7.6 meV and $2.2 \times 10^8/m$, respectively. Taking $\Sigma$ to be $8 \times 10^{12}/cm^2$ (ARPES value) [34], we estimated $E_F = 76$ meV and $k_F = 7.1 \times 10^8/m$.

A schematic diagram of the layered structure of the single-crystal Bi nanowire is shown in Figure 1.

Figure 2a shows the resistance of the nanowire as a function of the angle between the perpendicular to the Bi layers and the applied field. One observes a series of peaks in the magnetoresistance in the angular range between $\alpha = 20°$ and $100°$. The peaks observed as a function of angle, are Yamaji effect peaks, have been indexed with $n = 1, 2, 3$ and $4$. Fig 2b shows the linear dependence of the tangent of the angle with the index.

The electronic states of layered materials can be discussed, in first approximation, by the dispersion relation:

$$E = \frac{\hbar^2}{2m}(k_x^2 + k_y^2) - 2t \sin ck_z \qquad (1)$$

The first term represent the 2D surface states and the second part is the hopping term. Here $E$ is the energy, $m$ is the effective mass, $k_{x,y}$ is the x(y) component of the $k$ vector, $t_c$ is the interlayer energy and $c$ is the interlayer distance. Yamaji discovered that within the semiclassical approximation, in the extended zone, at certain "magic" angles the cross-section perpendicular to the magnetic field ascribes electronic orbits. The angles that Yamaji found are $ck_F \tan\alpha_n = (n-1/4)\pi$ where $n$ is an integer, $c$ is the interplanar distance and $k_F$ is the projection of the Fermi wave number on the layer conduction plane. Therefore, from the MR maxima n Fg. 2, we find the Fermi surface tube diameter. Taking $c = 0.38$ nm to be the interplanar distance we find $k_F = 1\times10^{10}$ m$^{-1}$. In comparison, taking $\Sigma$ to be $8 \times 10^{12}$/cm$^2$ (ARPES value, as ), $k_F$ can be estimated to be $7.1 \times 10^8$/m. Therefore there is a fair agreement. The structure of maxima and minima at $\alpha_n = 90°$, that is for B//C2, at high magnetic fields is reminiscent of the peak that is observed in layered materials and is attributed to coherent electron transport along small closed orbits on the sides of a cylindrical FS. The width of this peak allows estimation of the interlayer transfer integral $t_c$. [30, 31] Accordingly, $t_c = \Delta\alpha\, E_F/2k_F c$. This expression assumes a sinusoidal corrugation. The FWHM of the peak that is observed is $6°$. Assuming the values of the Fermi energy, 76 meV, and the Fermi wave length, $7.1 \times 10^8$/m in the review by Hofmann [34], one finds $t_c=3$ meV. The in-plane nearest neighbour hoppng energy is 3 eV and therefore the surface of bismuth, like the bulk is very anisotropic. The reason is the layered structure.

In contrast to the surface states where the FS is open, the bulk bismuth Fermi surface consist of ellipsoidal electron and hole pockets. The MR exhibit an angular dependency reflect this physics..[35]

Films of the well known topological insulator Bi(0.9)Sb(0.1) has been investigated [34]. The AMR shows the signature of layering but no evidence of a coherence peak.

We acknowlendge useful discussions with B. Halperin and R.Westervelt (Harvard) and wth Jeffrey Hunt (Boeing). This work was supported by the IIEN 15.817.02.09A, the NSF through Natonal Scence Foundation STC CIQM 1231319, the Boeing Company and the Keck Foundation.

**Conclusions**

Second order topologic order models of bismuth, predict an interlayer electrical coupling mediated by hinge states. Here we report that angle dependent magnetoresistance measurements of small diameter single-crystal bismuth nanowires exhibit the sequence of magnetoresistance (MR) peaks at Yamaji magic angles. The MR peaks are associated with magnetic field induced flat bands that arise because the semiclassical electron orbits are non dispersive for Yamaji angles. The MR also shows a peak for B//bilayer, indicating coherent transport between layers, and showing that the surface carriers Fermi surface is an open Fermi surface, a warped cylinder. This is in contrast to bulk Bi, that has closed Fermi surfaces for electrons and holes. Electrons do not loose coherence in the transfer between layers and therefore, as predicted, hinge states channel electrons without scattering across the layered structure of bismuth.

**Methods**

We fabricated 50-nm nanowires by applying Taylor's method, which involved stretching a wire to reduce its diameter. We started with fibres containing large-diameter (~200-nm) nanowires that can be fabricated using the Ulitovsky method. The bismuth is 99.999% pure. The fibres were stretched in a micropipette puller. In a microwire puller, a short section (a fraction of a millimetre in length) was brought to its softening point through the use of a coiled electrical filament that was coaxial with the fibre. When the softening point of the capillary tubing (~500°C) was reached, a mechanical pulling force was applied to each end of the fibre; the fibre deformed, and the heated section elongated and became thinner. Because the fibre contained a Bi filament, the action of the micropipette puller was to reduce the diameter of the glass fibre and the Bi nanowire simultaneously. In the third (and final) step of fabrication, the nanowires were placed in a travelling heater/oven in which a narrow region of a wire was heated, and this zone was moved slowly along the wire. Keeping in mind that Bi is easily oxidized in contact with air, it is expected that encapsulation of the Bi filament in the glass fibre protected our nanowire samples from oxidation. Electrical connections to the nanowires were performed using a $Ga_{0.5}In_{0.5}$ eutectic. The electrical and thermoelectric measurements were performed at the International Laboratory of High Magnetic Fields and Low Temperatures (ILHMFLT), Wroclaw, Poland. The maximum magnetic field was 14 T. The cryostat was equipped with a two angle rotator for handling the nanowire sample. The value of $R$ at low temperature ( 4 K) was $2.8 \times 10^{-6}\ \Omega^{-1}$.

FIGURE CAPTIONS

Figure 1. Schematic diagram of a single-crystal bismuth nanowire in a transverse magnetic field B , perpendicular to the wire axis, showing the crystalline bilayer structure. The angle dependent magnetoresistance is observed by rotating the sample around the wire, $\theta$ is the angular position of the nanowire. The angle between the perpendicular to the plane and the magnetic field is $\beta = \theta + 20° (90° - \theta)/90°$. The (111) atomic plane of the crystal is observed to be oriented at an angle of 70° from the wire axis in the nanowires employed in the present paper.

Figure 2. Yamaji oscillatons. (a) Angular dependence of the magnetoresistance for different magnetic fields $B$ in the plane perpendicular to the conducting plane (111), that is versus $\alpha$, of our 50-nm Bi nanowire sample. The wire is rotated as shown in figure 1. The resistance shows maxima at the magic angles indexed as n=1, 2, 3 and 4. The angles indicated as A, B and C are various minima of the resistance. For A, $\alpha = 54°$. For B, $\alpha = 57.4°$ and for C, $\alpha = 72°$. The corresponding magnetoresistance of the various cases of maxima and minima is studied in Figure 2.c showing saturated magnetoresistance for A, B and C . In large diameter nanowires, for $\alpha = 90°$ , $B//C2$. . Fig 2b shows the linear dependence of the tangent of the angle with the index.

Figure 3. Coherence peak. (a) Angular dependence of the magnetoresistance showing the coherence peak for $\alpha = 90°$. (b) Schematic diagram of the Fermi surface of surface charges in bsmuth. The clear circle shows the crosssection along the interlayer direction. The grey circle indicates the orbits that give rise to the coherence peak. The inset shows a schematic of the layer structure of the surface of bismuth.

........

Figure 1.

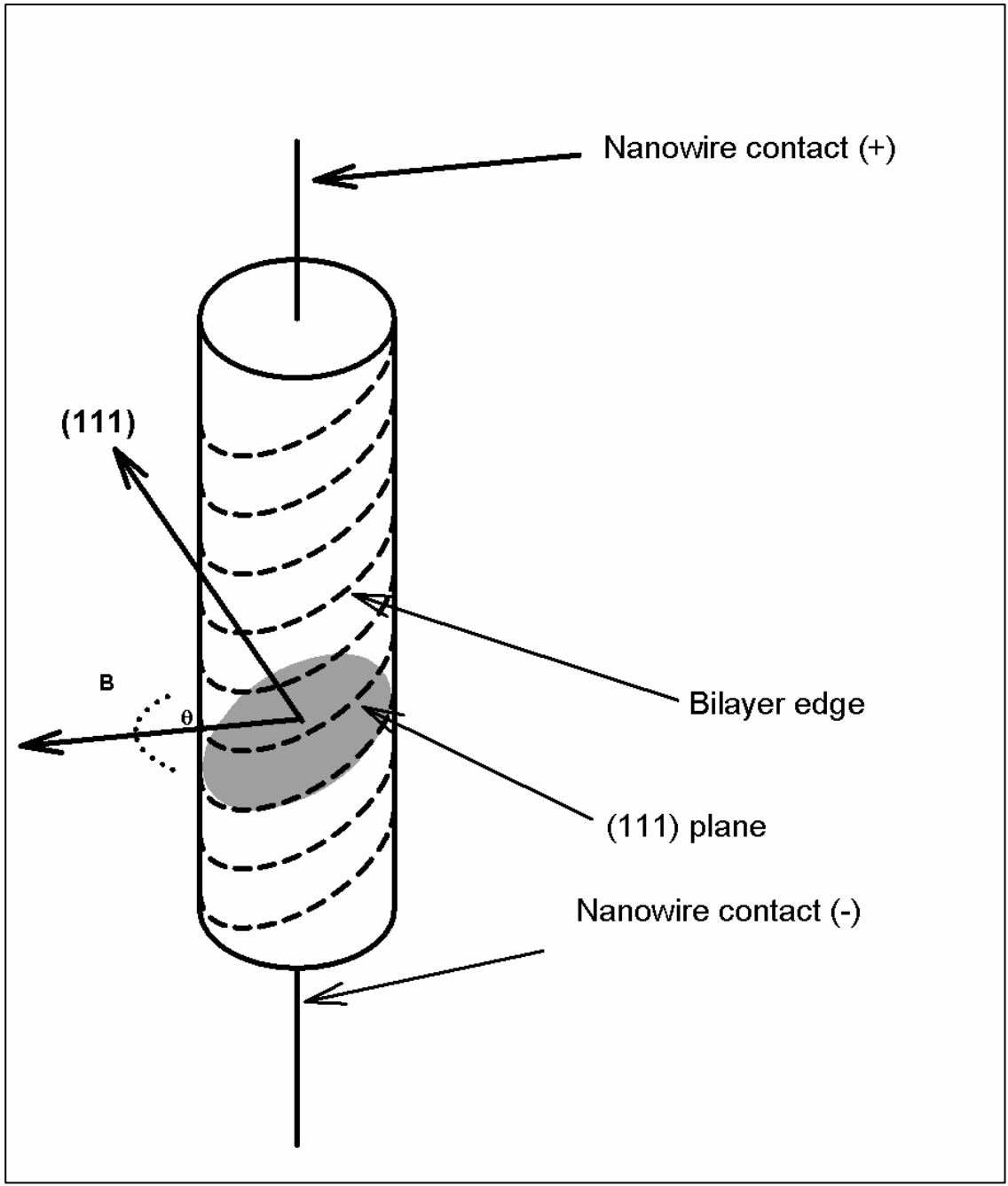

Figure 2a

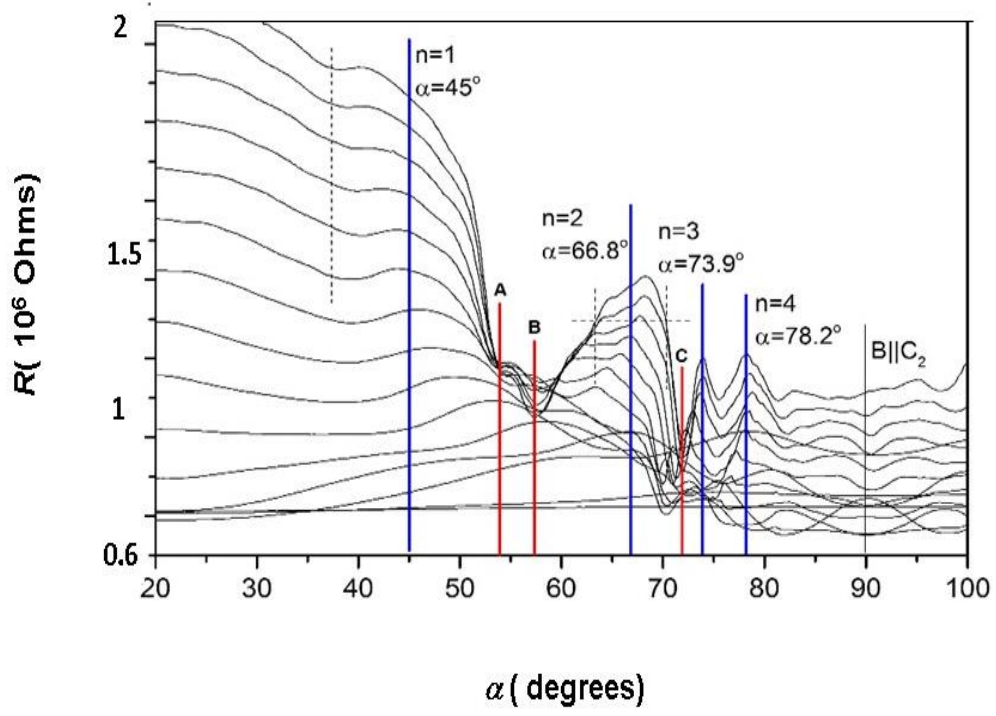

Figure 2 b. Index plot for Yamagi angles of Bi nanowires.

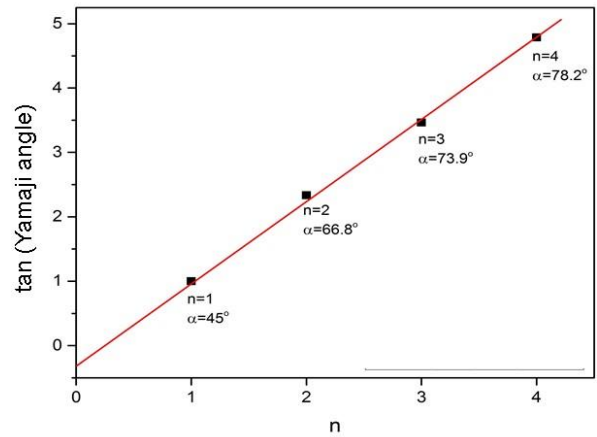

Figure 2 c. Anomalous magnetoresistance response of quantum confined 50-nm bismuth nanowires .

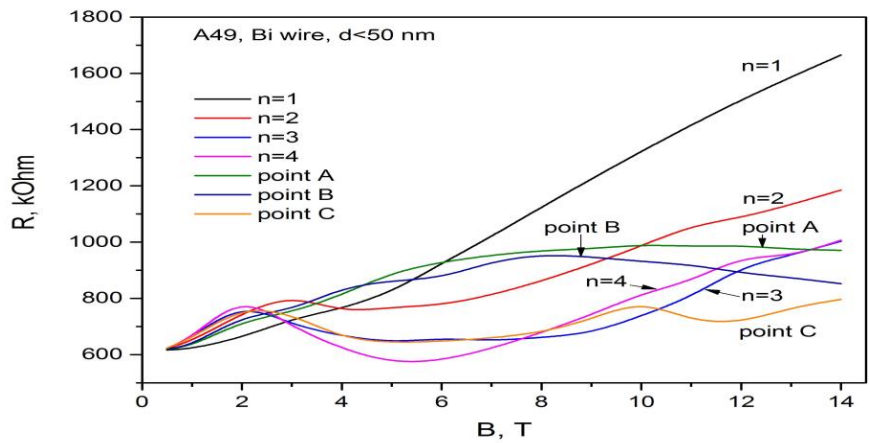

Figure 3a.

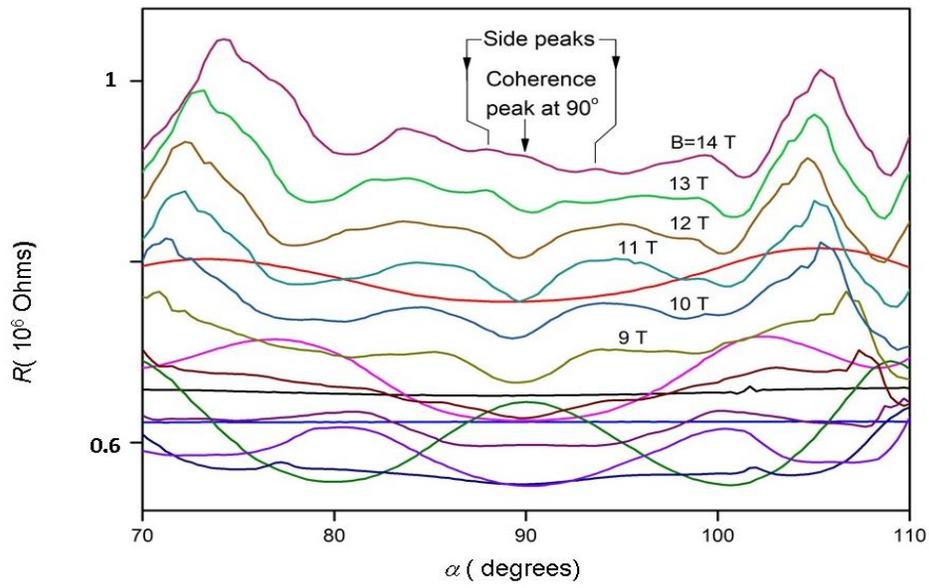

Figure 3b

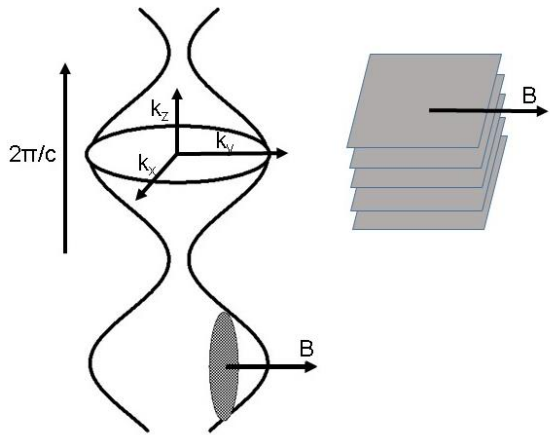